# An Improved Remote User Authentication Scheme Using Smart Cards

**Manoj Kumar**

Department of Mathematics, Rashtriya Kishan (P.G.) College

Shamli- Muzaffarnagar-247776

yamu_balyan@yahoo.co.in

*Abstract:* In 2000, Hwang and Li proposed a new remote user authentication scheme using smart cards. In the same year, Chan and Cheng pointed out that Hwang and Li's scheme is not secure against the masquerade attack. Further, in 2003, Shen, Lin and Hwang pointed out a different type of attack on Hwang and Li's scheme and presented a modified scheme to remove its security pitfalls. This paper presents an improved scheme which is secure against Chan-Cheng and all the extended attacks.

*Index Terms*: Authentication, cryptography, remote user, network security, password and smart card.

*Introduction:* Remote user password based authentication scheme, proposed by Lamport [14] in 1981, is a way to authenticate the remote user over an insecure channel. In Lamport's scheme, the *AS* stores a password table at the server to check the validity of the login request made by the user. However, high hash overhead and the necessity for password resetting decrease the suitability and practical ability of Lamport's scheme. Since then, many similar schemes [18]-[20] have been proposed. They all have a common feature: a verification password table should be securely stored in the *AS*. Actually, this property is a disadvantage for the security point of view. *U*nfortunately, the *AS* will be partially or totally braked/affected, if the password table is stolen /removed /modified by an antagonist/adversary.

Password authentication schemes with smart cards have a long history in the remote user authentication environment. So far different types of password authentication schemes with smarts cards [2]-[3]-[4]-[5]-[9]-[10]-[11]-[13]-[17]-[21]-[23] have been proposed. In 2000, Hwang and Li [16] proposed a new remote user authentication scheme using smart cards, which was based on ElGamal cryptosystem [22]. This scheme does not maintain the password table to check the validity of the login request. Also, it can withstand message-replaying attack. Chan and Cheng [7] pointed out the security weaknesses of Hwang and Li's Scheme and show that a legitimate user can easily create a valid pair of (*ID* , *PW* ) without knowing the secret key $x_s$ of the authentication server. In 2003, Shen-Lin- Hwang [11] discussed a different attack on the Hwang-Li's scheme and they also proposed a modified scheme to remove the security pitfalls of Hwang-Li's scheme.  In the same year, Chang and Hwang [1] explained the practical problems of the Chan – Cheng's attack on the Hwang-Li's scheme. Further, Leung - Cheng, - Fong and Chen [12] pointed out that the Shen-Lin-Hwang's scheme is still vulnerable to the attack proposed by Chan and Cheng.

*Contributions*: This paper presents an improved remote user authentication with smart cards. The improved scheme is secure against Chan – Cheng, Chan-Hwang and Shen-Lin- Hwang's extended attacks.

*Organization:* Section 2 reviews the Hwang – Li's scheme. Section 3 describes the cryptanalysis of Hwang – Li's scheme. Section 4 reviews the Shen-Lin-Hwang's scheme. Section 5 describes the security pitfall in the Shen-Lin-Hwang's scheme. An improved remote user authentication with smart card is proposed in section 6. Section 7 discusses the security of the improved scheme. Finally, comes to a conclusion in the section 8.

*Notations:* The notations used through out in this paper can be summarized as follows:

    *U* denotes the remote user.

    *ID* denotes the identity of the remote user *U*.

    *PW* denotes the password corresponding to the registered identity *ID*.

    *AS* denotes the authentication server.

    $x_s$ denotes the permanent secret key of the authentication server.

    f ( .) denotes a cryptographic one way hash function.

    $\leftrightarrow$ denotes a secret channel between the remote user *U* and the authentication server *AS*.

    $\rightarrow$ denotes a public channel (insecure channel) between the remote user *U* and the authentication server *AS*.

    *p* denotes a large prime number.

⊕ denotes XORed operation.

µ denotes a random number generated by the AS at the time of registration of the user U.

REVIEW OF THE HWANG-LI'S SCHEME: There are three phases in the Hwang-Li's scheme: the registration phase, login phase and the authentication phase. In the registration phase, the user U sends a request to the AS for the registration. The AS will issue a smart card and a password to every legal user through a secure channel. In the login Phase, when the user U wants to access the AS, she/he inserts her/his smart card to the smart card reader and then keys the identity and the password to access services. In the authentication phase, the AS checks the validity of the login request.

*Registration Phase:* A user U submits her/his ID to the AS. AS computes the password PW for the user U, as, $PW = ID^{x_s} \mod p$, where, $x_s$ is a secret key maintained by the AS and p is a large prime number. AS provides a password PW and a smart card to the user U through a secure channel. The smart card contains the public parameters (f, p), where f a one-way function.

*Login Phase:* User U attaches her/his smart card to the smart card reader and keys ID and PW. The smart card will perform the following operations:

Generate a random number r.

Compute $C_1 = ID^r \mod p$.

Compute $t = f(T \oplus PW) \mod p - 1$, where T is the current date and time of the smart card reader.

Compute $M = ID^t \bmod p$.

Compute $C_2 = M(PW)^r \bmod p$.

Sends a login request $C = (ID, C_1, C_2, T)$ to the AS.

*Authentication Phase:* Assume AS receives the message C at time $T_c$, where $T_c$ is the current date and time at AS. Then the AS takes the following action:

Check the format of *ID*. If the identity format is not correct, then AS will rejects this login request.

Check, whether $T_c - T \leq \Delta T$, where $\Delta T$ is the legal time interval due to transmission delay, if not, then rejects the login request C.

Check, if $C_2 (C_1^{x_s})^{-1} \stackrel{?}{=} (ID)^{f(T \oplus PW)} \bmod p$, then the AS accepts the login request. Otherwise, the login request will be rejected.

*Cryptanalysis of the Hwang- Li's Scheme:* This section discusses the security weakness of the Hwang – Li's scheme.

*Chan and Cheng's Attack: Impersonation Attack:* According to Chan and Cheng [7], a legal user Alice can easily generate a valid pair of identity and password without the knowledge of secrete key '$x_s$' of AS. Alice uses her valid pair ($ID_A$, $PW_A$) to generate another valid pair ($ID_B$, $PW_B$) as follows:

Alice computes $ID_B = (ID_A \times ID_A) \bmod p$. Then, she can compute the corresponding password $PW_B = ID_B^{x_s} \bmod p = (ID_A \times ID_A)^{x_s} \bmod p = (PW_A \times PW_A) \bmod p$. As a result, Alice can generate a valid pair ($ID_B$, $PW_B$) without knowing the secret key $x_s$.

*Shen-Lin-Hwang's Attack: Masquerading Attack:* According to Shen - Lin and Hwang [11] masquerading attack is possible on Hwang- Li's scheme. A user Bob can masquerade another user Alice to login a remote server and gain access right. Bob computes an identity $ID_B = ID_A^k \bmod p$, where $k$ is a random number such that gcd $(k, p)$ = 1. Then, he submits this identity $ID_B$ to *AS* for registration. *AS* provides a smart card and a password $PW_B = ID_B^{x_s} \bmod p$. With the knowledge of $PW_B$, Bob can compute the password $PW_A$ related to the identity $ID_A$, as, $PW_A = ID_A^{x_s} \bmod p = PW_B^{-k} \bmod p$. As a result, Bob can masquerade as Alice to login a remote server and gain access privilege.

*Chang- Hwang's Attack:* According to Chang and Hwang [1], there is a mistake in the Chan- Cheng's attack. It is not always possible that the square of a legal identity satisfies the specific identity format. Chang and Hwang generalized the Chan- Cheng's attack. They described two attacks.

*Attack- I:* Alice computes $ID_B = ID_A^k \bmod p$, where $k$ is a random number. Then, he can compute the corresponding password $PW_B = PW_A^k \bmod p$. As a result, a legal user Alice can impersonate other user Bob with a valid pair of *($ID_B$, $PW_B$)* to login the *AS.* If $ID_A$ is a primitive root of *Zp,* then all the valid identities and their corresponding password can be generated easily.

*Attack- II:* A group of eavesdroppers may cooperate to generate a valid pair of identity ($ID_G$, $PW_G$), as $ID_G = \prod ID_{Aj} \bmod p$ and $PW_G = \prod PW_{Aj} \bmod p$. Chang and Hwang pointed out that in Hwang – Li's scheme, it is still difficult to obtain the corresponding password for a known arbitrary valid identity, but once the valid identity is generated, its corresponding password will be obtained easily.

*Shen, Lin and Hwang's scheme:* Shen-Lin-Hwang [11] proposed a modified remote user authentication scheme to solve the security pitfall of the Hwang-Li's scheme. Shen-Lin-Hwang's scheme uses the concept of hiding the identity to prevent the masquerading attack. They modified the registration phase, now a shadow identity *SID* will be issued to the legal user. This modified in the registration phase is described below.

*Modified Registration Phase:* A user *U* submits her/his identity string *J* to the *AS* for the registration. The string *J* contains the name, address, unique number etc. This information in the string *J* is unique for every user. Then the *AS* computes a pair (*SID, PW*) for the user *U* after the identity *J is identified.* The pair (*SID, PW*) is computed as *SID* = *Red* (*J*) and *PW* = (*SID*)$^{x_s}$ mod *p*, where, *Red* (.) is a shadow identity of device which is only maintained in the remote server and *SID* is the shadow identity of the user *U*. Furthermore, the *AS* distributes the smart card and (*SID, PW*) to the user *U* in a secure way. The smart card contains the public parameters (*f, p*). In this scheme, the message sent to the *AS* now contains (*SID, $C_1$, $C_2$, T*). Because $J_i$ specially formatted, the evil user cannot compute new identity string $J_i$ via $SID_i$.

*Cryptanalysis of the Shen, Lin and Hwang's Scheme:* According to Leung-Cheng-Fong-Chan [12], Shen-Lin-Hwang's scheme [11] defends the attacks of registration for a new identity $ID_B$ via $ID_A$ for a legal user Alice. They also pointed out that the modified scheme is still vulnerable to the attack described by Chan and Cheng. They showed that the modified scheme is not secure against the attack that is similar to Chan - Cheng and Chang – Hwang's attacks. If we replace $ID_A$ with $SID_A$, then the Chang and Hwang's attack will work as follows. Alice computes $SID_B = (SID_A)^k \mod p$, where $k$ is a random number. Then, he can compute the corresponding password $PW_B = PW_A^k \mod p$. As a result, a legal user Alice can impersonate other user Bob with a valid pair of ($SID_B$, $PW_B$) to login the *AS*. If $SID_A$ is a primitive root of $Zp$, then all the valid identities and their corresponding password can be generated easily. Since, Chan – Cheng's attack is one case of this attack so it also works well.

*An Improved Scheme:* This section presents an improved remote user authentication scheme to remove the security pitfalls of the original scheme: Hwang – Li's scheme. This improved scheme overcomes the security flaws of Hwang – Li's scheme, which are described in the previous sections. The improved scheme has four phases: initial phase, registration phase, login phase and verification phase. These phases are described below.

*Initial Phase:* First, the insider at *AS* initialized the same public and secret parameters, as in Hwang –Li's scheme.

*Registration Phase:* For our requirement, we have modified the registration phase of Hwang and Li's scheme. Assume that this phase is executed over a secure channel. The following steps are involved in this phase.

Step R1. $U \leftrightarrow AS$: $J$. The string $J$ contains the name of the user $U$, address, identity $ID$ and a unique identification number etc, which are unique for the user $U$.

Step R2. $AS$ generates a random number $\mu$ and then computes $m = f(ID \oplus \mu) \bmod p$ and the password $PW$ for the user $U$, as, $PW = [f(ID \oplus \mu)]^{x_s} \bmod p$.

Step R3. $AS \leftrightarrow U$: a smart card containing the public parameters ($f$, $p$) and a pair ($ID \| \mu$, $PW$) to the user $U$.

After the registration, the identity $ID$ of the user has two parts: $ID$ and $\mu$. The second part $\mu$ is induced by the $AS$ after registration.

*Login Phase:* Whenever, the user $U$ wants to gain the access right on the $AS$, then the following steps are involved for the proper execution of this phase.

Step L1. $U$ attaches her/his smart card to the smart card reader at any time $T$ and keys her/his identity $ID \| \mu$ and the corresponding $PW$.

Step L2. The smart card of the user $U$ conducts the following computations:

Generate a random number $r$.

Computes $C_1 = f(ID \oplus \mu)^r \bmod p$.

Compute $t = f(T \oplus PW) \bmod p - 1$.

Compute $M = (ID)^t \mod p$.

Compute $C_2 = M(PW)^r \mod p$.

Step L3. $U \to AS$: $L_R = (ID \| \mu, C_1, C_2, T)$.

*Verification Phase:* Assume that the AS receives the login request $L_R$ at time $T_c$. Then, AS does the following computations to check the validity of the login request $L_R$.

Step V1. Check the specific format of the identity $ID \| \mu$. If the format of the identity is incorrect, then AS rejects the login request $L_R$.

Step V2. Check, whether $T_c - T \leq \Delta T$, where $\Delta T$ is the legal time interval due to transmission delay, if not, then AS rejects the login request $L_R$.

Step V3. Check, if $C_2 \stackrel{?}{=} (C_1^{x_s})(ID)^{f(T \oplus PW)} \mod p$, then the AS accepts the login request. Otherwise, the login request will be rejected by AS.

*Security Analysis of the Improved Scheme:* The above scheme is a modified form of the original scheme: *Hwang-Li's scheme*. The security analysis has been already discussed and demonstrated in [16]. Therefore, this section will only discuss the enhanced security features of the proposed scheme.

*Chan- Cheng's attack: Impersonation Attack:* Consider an antagonist/adversary, Alice wants to create another valid pair $(ID_B \| \mu_B, PW_B)$ such that $PW_B = [f(ID_B \oplus \mu_B)]^{x_s} \mod p$, by using her valid pair $(ID_A \| \mu_A, PW_A)$, which is provided by the AS to Alice through registration phase. In our scheme, this is infeasible to compute such a valid pair $(ID_B \| \mu_B, PW_B)$ due to the one-way property of hash function $f$.

Since, now the computed hash value $[f(ID_B \oplus \mu_B)]$ will not be equal to $[f(ID_A \oplus \mu_A)] \times [f(ID_A \oplus \mu_A)]$, then Chan – Cheng's attack will not work.

*Chang- Hwang's Attack:* Since, Chang- Hwang's Attack is an extended form of Chen- Cheng's attack, then this attack will also not work.

*Shen-Lin-Hwang's attack:* In Shen- Lin- Hwang's attack, suppose Bob is an adversary/attacker who wants to masquerade any valid user Alice. To create a favorable results for a successful attack, first the malicious user, Bob has to registered himself at AS. For this purpose, Bob intercepts the logon request $L_R = (ID_A \| \mu_A, C_1, C_2, T)$ from a public network and computes an identity $ID_B = ID_A^k \mod p$ or $ID_B = [f(ID_A \oplus \mu_A)]^k \mod p$, where k is a random number such that gcd (k, p) = 1. Then, he submits this identity $ID_B$ to AS for registration. Now, AS provides a smart card and a pair $(ID \| \mu_B, PW_B)$, where, $PW_B = [f(ID_B \oplus \mu_B)]^{x_s} \mod p$ to the antagonist user Bob. Now with the knowledge of $PW_B$, Bob will try to compute the password $PW_A$ related to the identity $ID_A$. On the one hand, If Bob submits the identity $ID_B = ID_A^k \mod p$, then he can try to computes the password $PW_A$ related to the identity $ID_A$, by using the relation:

$PW_A^* = PW_B^{-k} \mod p = [f(ID_B \oplus \mu_B)^{x_s}]^{-k} \mod p = [f(ID_B \oplus \mu_B)^{-k}]^{x_s} \mod p,$

$\neq [f(ID_A \oplus \mu_A)]^{x_s} \mod p \neq PW_A.$

Thus, the antagonist/attacker can compute only a fictious value $PW_A^* = [f(ID_B \oplus \mu_B)^{-k}]^{x_s} \mod p,$ which is different from the integral/authentic/existing password $PW_A = (ID_A \oplus \mu_A)^{x_s} \mod p$. On the other hand, If antagonist Bob submits the identity $ID_B = [f(ID_A \oplus \mu_A)]^k \mod p$, then also he can try to computes the password $PW_A$ related to the identity $ID_A$. With the similar reason, as described above, the antagonist Bob can not compute the integral/authentic/existing password $PW_A = (ID_A \oplus \mu_A)^{x_s} \mod p$. In this way, Bob is not able to compute the actual password $PW_A$ related to the identity $ID_A$. Thus the antagonist Bob can not masquerade as another user Alice to login a remote server. As a result, the improved scheme successfully withstands the masquerade attack: *Shen-Lin-Hwang's attack*

*Conclusion:* This paper proposed an improved remote user authentication scheme, which removes the security flaws of the original scheme: Hwang-Li's scheme. The proposed scheme is secure against all the attacks: Shen- Lin- Hwang's attack, Chan- Cheng's attack and Chang- Hwang's Attack.

**References**

1  CHANG C. C. and HWANG K. F.: 'Some forgery attack on a remote user authentication scheme using smart cards', Informatics, 2003, 14, (3), pp. 189 – 294

2  CHANG C. C. and HWANG S. J.: 'Using smart cards to authenticate remote passwords', Computers and Mathematics with applications, 1993, 26, (7), pp. 19-27 3  CHANG C. C. and WU T. C.: 'Remote password authentication with smart cards', IEE Proceedings-E, 1993, 138, (3), pp. 165-168

4  LEE C. C., LI L. H., and HWANG M. S.: 'A remote user authentication scheme using hash functions', ACM Operating Systems Review, 2002, 36, (4), pp. 23-29

5  LEE C. C., HWANG M. S., and YANG W. P.: 'A flexible remote user authentication scheme using smart cards', ACM Operating Systems Review, 2002, 36, (3), pp. 46-52

6  MITCHELL C. J., and CHEN l.: 'Comments on the S/KEY user authentication scheme', ACM Operating System Review, 1996, 30,( 4), pp. 12-16

7  CHAN C. K. and CHANG L. M.: 'Cryptanalysis of a remote user authentication scheme using smart cards', IEEE Trans. Consumer Electronic,2000, 46,(4), pp. 992-993

8  MITCHELL C.: 'Limitation of a challenge- response entity authentication', Electronic Letters, 1989, 25, (17), pp. 1195- 1196

9  SUN H. M.: 'An efficient remote user authentication scheme using smart cards', IEEE Trans. Consumer Electronic, 2000, 46, (4), pp. 958-961

10  CHIEN H. Y., JAN J. K., and TSENG Y. M.: 'An efficient and practical solution to remote authentication: smart card', Computer & Security, 2002, 21, (4), pp. 372-375

11  SHEN J. J., LIN C. W., and HWANG M. S.: 'A modified remote user authentication scheme using smart cards', IEEE Trans. Consumer Electronic, 2003, 49, (2), pp. 414-416 12 LEUNGK. C., CHENG L. M., FONG A. S., and CHEN and C. K.' 'Cryptanalysis of a remote user authentication scheme using smart cards', IEEE Trans. Consumer Electronic, 2003, 49, (3), pp. 1243-1245

13 LI L. H., LIN I. C., and HWANG M. S.: 'A remote password authentication scheme for multi-server architecture using neural networks', IEEE Trans. Neural Networks, 2001, 12, (6), pp. 1498-1504

14 LAMPORT L.: 'Password authentication with insecure communication', communication of the ACM, 1981, 24 (11), pp. 770-772

15 KUMAR M.: 'Some remarks on a remote user authentication scheme using smart cards with forward secrecy', IEEE Trans. Consumer Electronic, 2004, 50 (2), pp. 615-618

16 HWANG M. S., and LI L. H.: 'A new remote user authentication scheme using smart cards', IEEE Trans. Consumer Electronic, 2000, 46, (1), pp. 28-30

17 UDI M.: 'A simple scheme to make passwords based on the one-way function much harder to crack', Computer and Security, 1996, 15, (2), pp. 171 - 176

18 LENNON R. E., MATYAS S. M., and MAYER C. H. 'Cryptographic authentication of time-variant quantities', IEEE Trans. on Commun., 1981, COM - 29, (6) , pp. 773 - 777

19 Wang S. J.: 'Yet another login authentication using N-dimensional construction based on circle property', IEEE Trans. Consumer Electronic, 2003, 49, (2), pp. 337-341 20  YEN S. M., and LIAO K.H.: 'Shared authentication token secure against replay and weak key attack', Information Processing Letters, 1997,pp. 78-80

21  WU T. C.: 'Remote login authentication scheme based on a geometric approach', Computer Communication,1995, 18, (12), pp. 959 - 963

22  ELGAMAL T.: 'A public key cryptosystem and a signature scheme based on discrete logarithms', IEEE Trans. on Information Theory, 1985, 31, (4), pp. 469-472

23  TANG Y. L., HWANG M. S. and LEE C. C.: 'A simple remote user authentication scheme', Mathematical and Computer Modeling, 2002, 36, pp. 103 – 107

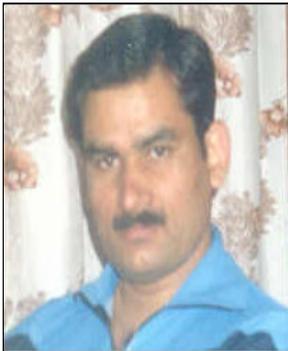

**Manoj Kumar** received the B.Sc. degree in mathematics from Meerut University Meerut, in 1993; the M. Sc. in Mathematics (Goldmedalist) from C.C.S.University Meerut, in 1995; the M.Phil. (Goldmedalist) in *Cryptography*, from Dr. B. R. A. University Agra, in 1996; the Ph.D. in *Cryptography*, in 2003. He also qualified the *National Eligibility Test* (NET), conducted by *Council of Scientific and Industrial Research* (CSIR), New Delhi- India, in 2000.

He also taught applied Mathematics at D. A. V. College, Muzaffarnagar, India from Sep 1999 to March 2001; at S.D. College of Engineering & Technology, Muzaffarnagar- U.P. – INDIA from March 2001 to Nov 2001; at Hindustan College of Science & Technology, Farah, Mathura- U.P. – INDIA, from Nov 2001 to March 2005. In 2004, the Higher Education Commission of U.P. has selected him. Presently, he is working as lecturer in Department of Mathematics, R. K. Inter College Shamli- Muzaffarnagar- U.P. – INDIA.

He is a member of Indian Mathematical Society, Indian Society of Mathematics and Mathematical Science, Ramanujan Mathematical society, and Cryptography Research Society of India. His current research interests include Cryptography, Numerical analysis, Pure and Applied Mathematics.